\documentclass{aip-cp}

\usepackage[numbers]{natbib}
\usepackage{rotating}
\usepackage{graphicx}

\begin{document}

\title{An Analysis of the New LHC Data \\ through the Dispersion Relations}

\author[aff1]{O.V. Selyugin\corref{cor1}}
\author[aff2]{J.-R. Cudell\corref{cor2}}

\affil[aff1]{BLTPh, JINR, Dubna,Russia}
\affil[aff2]{STAR Institute, University of Li\`ege, Belgium}
\corresp[cor1]{Corresponding author: selugin@theor.jinr.ru}
\eaddress{jr.cudell@ulg.ac.be}
\maketitle
\begin{abstract}
 We present an analysis of the new experimental data obtained by the TOTEM  and ATLAS Collaborations at the LHC at $\sqrt{s} = 7 $  and $8
 \ $TeV and at small momentum transfer.
    We analyse the tension between the (indirect) measurements of the total cross section, and show
the impact of various assumptions on the extraction of the parameters from the elastic scattering amplitude, with a special attention to   the total
cross section. In particular, the determination of the phase of the elastic scattering amplitude will play an important role, and we shall study it
via dispersion relations. We shall also examine the origin of the dependence on momentum transfer of the slopes of the different parts of the
scattering amplitude in different models.
We shall also give the results of another  similar analysis based on a Regge-trajectory approach for the hadron scattering amplitude.
\end{abstract}
\section{INTRODUCTION}
  The measurement of the $s$-dependence of the total cross sections $\sigma_{tot}(s)$
   and of $\rho(s,t)$ $-$ the ratio of the real part to the imaginary part of the
 elastic scattering amplitude $-$  is very important \cite{Rev-LHC} as it is a test of the first principles of quantum field theory.
 These principles lead to the integral dispersion relations that relate the real and the imaginary parts of the elastic scattering amplitude
  \begin{eqnarray}
  && \rho_{ \frac{pp}{p\bar{p}} }(s,0)\  \sigma_{ \frac{pp}{p\bar{p}} }(s) =   \frac{A_{\frac{pp}{p\bar{p}}}(m^2)}{p}
     + \frac{E}{\pi p}  \int_{m}^{\infty} dE^{'}p^{'}\left[\frac{ \sigma_{ \frac{pp}{p\bar{p}} }(E') }{E'(E'-E)} -
     \frac{ \sigma_{ \frac{p\bar{p}}{pp} }(E') }{E'(E'+E)}\right].
   \end{eqnarray}
  where $E$ is the fixed-target energy, i.e. $s=2 m_p(m_p+E)$. Hence, in theory, the scattering amplitude
    has to satisfy analyticity in the Mandelstam representation, and the real part of the scattering amplitude must be derivable from
    the imaginary part \cite{Roy}.

 For simplicity one often uses the local or derivative dispersion
 relations (see for example \cite{Bronzan,GMS})    to determine the real part of the scattering amplitude.
     For example the COMPETE Collaboration used, for $C=+1$ part of the amplitude \cite{COMPETE}:
 \begin{equation}
 Re A_{+}(s,0) / \left(\frac{E}{m_{p}}\right)^{\alpha} =\tan\left[\frac{\pi}{2} \left(\alpha-1+E\frac{d}{dE}\right) \right]\left[
  Im A_{+}(s,0)/(\frac{E}{m_{p}})^{\alpha}\right].
   \end{equation}
   A different form for the derivative dispersion relation can also be found  \cite{Roy-DR}
    \begin{equation}
 Re A_{+}(s,0)/ImA_{+}(s,0)	 = \left(\frac{\pi}{\log(s/s_{0})}\right) \frac{d}{d\tau}[\tau Im A_{+}(s,t)/Im_{+}(s,t=0)] ,
 \label{Roy-DR}
   \end{equation}
   where $ \tau= t\ \log^2(s/s_{0})$ and $s \rightarrow \infty$.
     To satisfy these relations the scattering amplitude must be
      a unique analytic function of its kinematic variables, and connect different reaction channels.

A very precise measurement of $\rho(s,t)$ at the LHC would give the possibility to check the validity of the dispersion relations
\cite{Rev-LHC}.   In turn, as indicated in  \cite{Khuri1,Khuri2}, a deviation in the phase of the scattering amplitude could result from the
existence of a fundamental length.

\section{THE DIFFERENTIAL CROSS SECTION AT SMALL MOMENTUM TRANSFER}
 Now the TOTEM and ATLAS Collaborations have already produced five sets of data at small momentum transfer at
 $7$ TeV and $8$ TeV  (see Table 1).
\begin{table*}[h]
\caption{The  LHC elastic scattering data at small $t$ at $7$ and $8$ TeV.}
\label{tab:a}
\begin{center}
\tabcolsep12pt\begin{tabular}{|c|c|c|l|l|c|c|}
\hline
$\sqrt{s}$ [TeV]  & Collabor. & N-points & $t_{min}$ [GeV$^2$] & $t_{max}$ [GeV$^2$]   & ref. & date \\  \hline
$7$   & TOTEM & 82 & $0.00515$ & $0.371$    & \cite{T7a} & 17.08.2012  \\
$7$  & ATLAS & 40 & $0.0062$ & $0.3636$  & \cite{ATL7}  &  25.08.2014 \\
\hline
$8$    & TOTEM$_{a}$ & 30 & $0.0285$  & $0.1947$ &  \cite{T8a} & 12.09.2015   \\
$8$   & TOTEM$_{b}$ &$31$ & $0.000741$ & $0.201$  & \cite{T8b}  & 11.12.2015  \\
$8$    & ATLAS  & $39$  & $0.0105$ & $0.3635$  & \cite{ATL8}   & 25.06.2016  \\
\hline
\end{tabular}
\end{center}
\end{table*}

   From these new data, the value of the total cross sections was extracted via different methods. At $7$ TeV, the TOTEM Collaboration
   obtained four values for $\sigma_{tot}$ (see Table 2).
   \begin{table}[h]
 \caption{The values of the total cross section  $\sigma_{tot}(s)$, mb are obtained at LHC  and in the HEGS model}
\label{Table-3}
\begin{tabular}{|c|c|c|c||c|} \hline
  $\sqrt{s}$ [GeV]    & $\sigma_{tot}^{TOTEM}(s)$ [mb]  &  $\sigma_{tot}^{ATLAS(s)}$ [mb]  &
  $\sigma^{TOTEM}_{tot}- \sigma^{ATLAS}_{tot}$ [mb]& $\sigma_{tot}(s)^{HEGS}$ [mb]   \\  \hline
       & $98.3 \pm 2.8 $ &     &   &  \\
       & $98.6 \pm 2.2$  &     &  &   \\
$7000$ & $99.1 \pm 4.3$   &     &  &    \\
       & $98.0 \pm 2.5$  &     &  & \\
       & combined=$98.5 \pm 2.9$  & $95.35 \pm 2.0$   & $3.15$  & $97.4 $   \\   \hline
 $8000$ & $101.7 \pm 2.9$   & $96.07 \pm 1.34$   &$5.6$   & $ 98.9$        \\  \hline
\end{tabular}
\end{table}
These data are consistent and their mean value is equal $98.5$ mb.
The ATLAS Collaboration, using their  differential cross section data in a region of $t$ where the Coulomb-hadron interference is
negligible, obtains the value $\sigma_{tot}=95.35 \pm 2.0$ mb. The difference between the two results, $\sigma_{tot}(s)$(T.) -
$\sigma_{tot}(s)$(A.) = $3.15$ mb, is about  1 $\sigma$.
 At $8$ TeV, the measured value of $\sigma_{tot}$ grows, especially in the case of the TOTEM
 Collaboration (see Table 2) and the difference between the results of the two collaborations grows to
$\Delta(\sigma_{tot}(s)$(T.) - $\sigma_{tot}(s)$(A.) $= 5.6$ mb, i.e. 1.9 $\sigma$. This is reminiscent of
 the old situation with the measurement of the
 total cross sections at the Tevatron at $\sqrt{s} = 1.8$~TeV via the luminosity-independent method
by different collaborations.

To compare these different sets of the data we need in a gauge.
 Although the value of the total cross section was expected
\cite{COMPETE}, the first data obtained at the LHC at $7$ TeV by the TOTEM Collaboration \cite{T7a}
were at odds with all predictions.  One of us developed a new model,
 High Energy Generalized Structure (HEGS) \cite{HEGS0},
 that describes well  all high energy data on  elastic proton-proton and proton-antiproton scattering with only a few free parameters.
This model was further developed  \cite{HEGS1} to describe quantitatively
 the data in the wide energy interval $9.8$ GeV $\leq \sqrt{s} \leq 8.0 $ TeV
 and in the wide region of momentum transfer $0.000375 \leq |t| \leq 15 $ GeV$^2$, at the cost of a few low-energy free parameters.

HEGS assumes a Born term for the scattering amplitude which gets unitarized via
the standard eikonal representation to obtain the full
elastic scattering amplitude. The scattering amplitude has exact $s\leftrightarrow u$ crossing symmetry as it is written
in terms of a complexified Mandelstam variable $\bar{s} = s e^{-i\pi/2}$ and this determines its real part.
The scattering amplitude also satisfies  the integral dispersion relation at large $s$.
It can be thought of as the simplest unified analytic function of its kinematic variables connecting different reaction channels
without additional terms for separate regions of momentum transfer or energy.
 Note that it reproduces the diffraction minimum of the differential cross section
 in a wide energy region \cite{HEGS-dm}.
HEGS describes the experimental data at low momentum transfer, including the Coulomb-hadron interference region, and hence it
 includes all five electromagnetic spin amplitudes and the Coulomb-hadron interference phase.

 Let us compare the predictions of the HEGS model for the differential
 elastic cross section at small $t$ with the LHC data. In the fitting procedure only the statistical errors are taken
  into account. The systematic errors are reflected through an additional normalization coefficient which is the same
  for all the data of a given set. The different normalization coefficients
  have practically random distributions at small $t$ (see the Tables in \cite{HEGS1}).
 In the present case, we fix all the parameters of the model but the normalization
  coefficient.  The model then reproduces well all the data sets but the normalization coefficients
  are somewhat different for the data of TOTEM and ATLAS (see Table 3).
 \begin{table*}[h]
\caption{A  comparison of the  data from TOTEM and ATLAS with the HEGS model at $\sqrt{s_1}=7$ and $\sqrt{s_2}=8$ TeV.}
\label{tab:c}
\begin{center}
\tabcolsep12pt\begin{tabular}{|c|c|l|l|c|}
\hline
    Collaboration    &\multicolumn{3}{|c|}{ normalization}        & $[\sigma_{tot}(s_2)- \sigma_{tot}(s_1)]$  [mb]   \\  \hline
            &$k_{s_1}$  &    $k^{a}_{s_2}$  &  $k^{b}_{s_2}$       &  \\  \hline
 TOTEM & $0.94$   & $0.91$ & $0.9 $  & $ 3.2$ \\
 ATLAS &  $1.0$  & $1.0$ & $ $    & $1.15$ \\     
HEGS   & $1.0$ & $1.0$  & $ $ &  $ 1.5$    \\
\hline
\end{tabular}
\end{center}
\end{table*}
Of course, we cannot say that the normalization of the ATLAS data is better than that of the TOTEM data simply because it coincides with the
HEGS predictions. But this exercise may point to the main reason for the different values of the total cross section obtained by  the
two collaborations.

This does not exclude some further problems with the analysis
of the experimental data, e.g. those related to the analysis of the TOTEM data at $\sqrt{s}=7$ TeV \cite{Sel-NP1}.
First of all,  the behavior of the real part of the scattering amplitude is usually taken
 as proportional to the imaginary part. Hence the slopes of both parts are equal.
 Secondly,  some unusual assumption about the growth of the real part of the scattering amplitude at small momentum transfer
  (the so-called ``peripheral case" \cite{T8a}). Both assumptions violate analyticity as they do not respect the dispersion relations.
The latter lead to a slope for the real part of the scattering amplitude
 larger than the slope of its imaginary part.
To see this, imagine that the imaginary part of the elastic scattering amplitude takes a simple exponential form $Im A_{+} \sim h e^{Bt}$ then
from Equation (\ref{Roy-DR}) one obtains
 $ ReA_{+} = (1.+Bt)e^{Bt}$, which has a zero in the
 region of momentum transfer around $|t| \sim 0.1 - 0.15$ GeV$^2$, and hence falls faster than the imaginary part.
 Note that any unitarization procedure will enhance this difference.

 This also shows that the differential cross section at small $|t|$ should not fall as a simple exponential.
 This was announced as a discovery in \cite{T8a} although such a behavior  was
 noted a long time ago  at lower energy. Most dynamical models that describe elastic scattering also lead
 to a non-exponential behavior of the differential cross section at small $t$.
For example, the  Dubna Dynamical Model  \cite{DDM},  which takes into account the contribution from the meson cloud of the nucleon  and
uses the standard eikonal form of the unitarization, leads to a Born term for the scattering amplitude in the impact parameter representation of
the form $ h s^{\Delta} exp[ -\mu(s) \sqrt{b_{0}^{2}+b^{2}}] $. After unitarization, the slope of the scattering amplitude becomes non linear in $t$
as it contains the term $ b_{0} \sqrt{\mu^{2}-t}$. Such a behavior was obtained in many works \cite{sqr-N} and is  based on the inclusion
of the  two-pion threshold  \cite{G-P-62}.
\begin{table*}[h]
\caption{The non-exponential behavior of the differential cross section at $\sqrt{s}=540$ GeV and $\sqrt{s}=1800$ GeV. }
\label{tab:d}
\begin{center}
\begin{tabular}{|c|c|c||c|c|}
\hline
& \multicolumn{2}{|c||}{ $\sqrt{s}=540$ GeV } &\multicolumn{2}{|c|}{$1800$ GeV} \\  \hline
{$-t$ [GeV$^2$]} &
{$\rho(s,t)$}   &
 { $B(s,t)$ [GeV$^{-2}$] } &
 {$\rho(s,t)$}   &
 {$B(s,t)$ [GeV$^{-2}$]} \\
\hline
$0.001$  & 0.141 & 16.8 & 0.182 & 18.1 \\
$0.014$  & 0.135 & 16.5 & 0.178 & 17.7 \\
$0.066$  & 0.112 & 15.5 & 0.161 & 16.6 \\
$0.120$  & 0.089 & 14.9 & 0.143 & 15.9 \\
\hline
\end{tabular}
\end{center}
\end{table*}

An analysis of the high-energy data for proton-antiproton scattering in the framework of this model
 shows an obviously non-exponential behavior of the differential cross sections   (see Table 4).
 For $\sqrt{s} = 540 $ GeV, the slope changes
 from $16.8$ GeV$^{-2}$ to $14.9$ GeV$^{-2}$ as one goes from $t=-0.001$ GeV$^2$  to $t=-0.12$ GeV$^2$.
 For the same $t$ interval,  $\rho(s,t)$ changes from $ 0.141$ up to $ 0.089$. Similarly, at
 $\sqrt{s} = 1800 $ GeV, the slope changes  from $18.1$ GeV$^{-2}$ to $15.9$ GeV$^{-2}$ as one goes from $t=-0.001$ GeV$^2$ to
 $t=-0.12$ GeV$^2$, and  $\rho(s,t)$ changes, this time  from $ 0.182$ up to $ 0.143$.
 Hence the model shows a continuous decrease of the slope and $\rho$ at small $t$.
Similar results were obtained in \cite{Pumplin-Sl} in the framework of another eikonalized model.
This is not the place for a careful explanation of all the features of this non-exponential behavior,
which we shall postpone to a future publication.
\section{FIT TO THE TOTAL CROSS SECTION AT LHC ENERGIES}
One can study the LHC  data simultaneously and fit them to a simple function,
paying special attention to the normalization. We take the scattering amplitude
as used by the TOTEM Collaboration \cite{T8b},
 where the slope of the imaginary and real parts of the scattering amplitude was determined by three terms
 $ (B_{1}+B_{2} t +B_{3} t^{2})/2$. We introduce a $\log(s)$ dependence for the slope and
 $\log^2(s)$ dependence for $\sigma_{tot}(s)$, and take
 a constant value for $\rho$.
\begin{equation}
A(s,t)/s=h~(i+\rho)\log^2(s)~\sigma_{tot} (\sqrt{s})^{B_1+B_2~t+B_3~t^2}\label{eq4}
\end{equation}
 In this case we have 5 free parameters. We include only
statistical errors and the systematical errors are reflected
in the additional data-normalization coefficients $k_{i}$. If these coefficients
   are fixed to $1$ then the  $\chi^2$ is enormous  and the value of   $\sigma_{tot}(s)$
   is closer to that of ATLAS data then to that of TOTEM  (see Table 5 first column).

     If the normalization coefficients are taken as free parameters, except for the last ATLAS data at
     $\sqrt{s}=8 $ TeV, the $\chi^2$ decreases substantially and the value of   $\sigma_{tot}(s)$
     decreases by $1 $ mb (and is very close to the ATLAS value).
     If all coefficients are free, the $\chi^2$ decreases and $\sigma_{tot}(s)$
     increases above both the ATLAS and the TOTEM values.
    Note that the ratio of the normalization coefficients remains practically the same, with the TOTEM data  above the ATLAS data by about
    $10\% $.

      Now let us examine the case where the real part of the scattering amplitude is determined by
      the complex $\bar{s}$ as required by crossing symmetry. The power of $s$ will then be taken
      in the form
      \begin{equation}
      \alpha'(t)=\alpha_{1}^{\prime} \ t  + D (\sqrt{4 m_{pi}^2-t}- 2 m_{pi}).\label{eq5}
      \end{equation}
\begin{table*}[h]
\caption{The fit of the sum of the five sets of the LHC data}
\label{tab:a}
\begin{center}
\tabcolsep12pt\begin{tabular}{|c||c|c|c|c|c|}
\hline
&\multicolumn{3}{c|}{Equation \ref{eq4}}&\multicolumn{2}{c|}{Equation \ref{eq5}}\\ \hline
 $k_{TOTEM-7~\rm TeV}$ & $1$  & $0.93$ & $1.14$ &1  &0.93 \\  \hline
 $k_{ATLAS-7~\rm TeV}$ & $1$  & $0.98$ & $1.18$ &1  &0.98\\  \hline
 $k_{TOTEM-8~\rm TeV (a)}$ & $1$  & $0.98$ & 1.1 &1&0.9\\  \hline
 $k_{TOTEM-8~\rm TeV (b)}$ & $1$  & $0.9$ &  1.1&1&0.901\\  \hline
 $k_{ATLAS-8~\rm TeV}$ & $1$  & $1$ & $1.2$  &1& 1.02\\  \hline
$ \chi^2 $               & 48212&2872&1508&4774&1327\\ \hline
 $\sigma_{tot}$(7 TeV) [mb]&96.3&95.3&106.1&96.1&95.7\\ \hline
  $\sigma_{tot}$(8 TeV) [mb]&99.2&98.2&109.3&99.0&98.6\\ \hline

\end{tabular}
\end{center}
\end{table*}
In this case only three parameters are fitted (plus the five normalization
coefficients).  If all $k_{i}$ are fixed at $1$,  the values of  $\sigma_{tot}(s)$ are similar to the
 previous case but the value of $\chi^2$ decreases ten times.  If the $k_{i}$ are fitted
 (and bounded by $0.9 \leq k_{i} \leq 1.1$) then the $\chi^2$ has a minimal value and
  the values of $\sigma_{tot}(s)$ are similar again to the previous case.
\section{CONCLUSION}
The new data on  $\sigma_{tot}(s)$ obtained by the TOTEM and ATLAS Collaborations
at $\sqrt{s} = 8 $  TeV differ by 6\%.   Our analysis  of the new data  on elastic $pp$ scattering
at small $t$ and at $\sqrt{s} = 7 $ and $8$ TeV  is
based on the crossing symmetry which the scattering amplitude must satisfy (and which invalidates
the ``peripheral case"  used in  \cite{T8b}).
The new High Energy Generalized Structure model (HEGS), based on these analytic properties,
gives a good  description of all the elastic nucleon scattering amplitudes at high energy  with only
6 parameters.

The HEGS model suggests that the discrepancy comes from the normalization of the TOTEM and ATLAS data.
A purely phenomenological analysis (Table 5) gives the same results.
Our analysis leads to values of $\sigma_{tot}(s)$ slightly above the ATLAS value, and significantly below
 the TOTEM result.

 Of course a more careful examination of the detailed structure of the slope  $ B(s,t)$  (non-exponential, oscillating)
 and of the impact of  the unitarization procedure is needed. In particular, one needs to take
 into account the form factors of hadrons and that fact that the slope of $Re A(s,t)$ exceeds that of $Im A(s,t)$.

  The problems with the normalization could be solved via a measurement of the elastic cross sections in the deep
  Coulomb-hadron interference region.
  Hence we need  new high-precision data at small $|t|$ at $13$ TeV.


\section{ACKNOWLEDGMENTS}
O.V.S. would like to thank the organizers R. Fiore and A. Papa for the invitation
and support of his participation in the conference. Part of this research has benefitted
from a visiting grant from FRNS.
\bibliographystyle{aipnum-cp}

\end{document}